\documentclass[aps,prl,twocolumn,showpacs,amsmath,amssymb,preprintnumbers,10pt]{revtex4-1}
\usepackage{blindtext}
\usepackage{epsfig}
\usepackage{amsfonts}
\usepackage{amssymb}
\usepackage{enumitem}
\usepackage{amsthm}
\usepackage{subfig}
\usepackage{bm}
\usepackage{hyperref}
\usepackage{mathrsfs}
\usepackage{bbm}
\usepackage{cancel}
\usepackage[dvipsnames]{xcolor}
\usepackage{accents}
\usepackage{relsize}
\usepackage{float, slashed, graphicx, amssymb, amsmath}
\usepackage{shuffle}
\usepackage{tikz}
\usetikzlibrary{hobby,calc,shapes, positioning, backgrounds,trees, decorations, decorations.markings, decorations.pathmorphing, arrows, shapes.geometric, snakes}
\usepackage{tikz-feynman}
\tikzfeynmanset{compat=1.1.0}
\tikzfeynmanset{
graviton/.style={circle, draw=green!60, fill=green!5, very thick, minimum size=7mm}
}
\tikzfeynmanset{
codot/.style={/tikz/shape=circle,
/tikz/fill=red,/tikz/minimum size=0.1cm,/tikz/inner sep=1.8pt}
}
\tikzfeynmanset{sb/.style=
{/tikz/shape=circle,
/tikz/fill=black,
 /tikz/minimum size=0.3cm,/tikz/inner sep=1.pt
 } }
\tikzfeynmanset{myblob/.style=
{/tikz/shape=rectangle,
/tikz/fill=red,
 /tikz/minimum size=0.2cm,
 } }
\tikzset{box/.pic={\filldraw[fill=black]  (0,0) circle (2.5pt);
				   \filldraw [fill=black] (0.5,0) circle (2.5pt);
			       \draw [line width=5pt] (0,0) -- (0.5,0);}}

\usepackage[T1]{fontenc} 

\makeatletter
\newcommand \UPlus {\mathop {\operator@font \uplus }\limits }
\makeatother
\makeatletter
\newcommand \Bigcup {\mathop {\operator@font \bigcup }\limits }
\makeatother
  \def\LabelNote#1{}
 \def\Label#1{\label{#1}%
  \smash{\hbox to0pt{\raise1ex\hbox{\tiny[#1]}\hss}}}
  
  \def\Cdot{{\cdot}}
  
\def\nn{\nonumber}

\newcommand{\red}{\color{red}}

\newcommand{\white}{\color{white}}

\def\spa#1.#2{\left\langle#1\,#2\right\rangle}
\def\spb#1.#2{\left[#1\,#2\right]}

\def\be{\begin{equation}}
\def\ee{\end{equation}}
\def\bea{\begin{eqnarray}}
\def\eea{\end{eqnarray}}  
\allowdisplaybreaks

\newcommand{\npre}{\mathcal{N}}  

\newcommand{\commut}{\Gamma}
 
\begin{document}

\preprint{
NORDITA 2021-091,
	QMUL-PH-21-45,
	SAGEX-21-34, 
	UUITP-60/21
}

\title{
Kinematic Hopf Algebra for BCJ Numerators in \\ Heavy-Mass Effective Field Theory and Yang--Mills Theory}
\author{Andreas Brandhuber$^1$}
\email{a.brandhuber@qmul.ac.uk}
\author{Gang Chen$^1$}
\email{g.chen@qmul.ac.uk}
\author{Henrik Johansson$^{2,3}$}
\email{henrik.johansson@physics.uu.se}
\author{Gabriele Travaglini$^1$}
\email{g.travaglini@qmul.ac.uk}
\author{Congkao Wen$^1$}
\email{c.wen@qmul.ac.uk}
\affiliation{$\mbox{}^{1}$Centre for Theoretical Physics, Department of Physics and Astronomy, 
Queen Mary University of London,\\ Mile End Road, London E1 4NS, United Kingdom}
\affiliation{
$\mbox{}^{2}$Department of Physics and Astronomy, Uppsala University, Box 516,  75120 Uppsala, Sweden}
\vspace{0.9cm}
\affiliation{$\mbox{}^{3}$Nordita, Stockholm University and KTH Royal Institute of Technology,  Hannes Alfv\'{e}ns  v\"{a}g 12, 10691 Stockholm, Sweden}

\begin{abstract}
We present a closed formula for all Bern-Carrasco-Johansson (BCJ) numerators describing  $D$-dimensional tree-level scattering amplitudes in a heavy-mass effective field theory with two  massive particles and an arbitrary number of gluons. The corresponding gravitational amplitudes obtained via the double copy directly enter the computation of black-hole scattering and gravitational-wave emission. Our construction is based on finding a kinematic algebra for the numerators, which we relate to a quasi-shuffle Hopf algebra.  The BCJ numerators thus obtained have a compact form and intriguing features: gauge invariance is manifest, locality is respected for massless exchange, and they  contain poles corresponding to massive exchange. 
Counting the number of terms in a BCJ numerator for $n{-}2$ gluons gives the Fubini numbers $\mathsf{F}_{n-3}$, reflecting the underlying quasi-shuffle Hopf algebra structure. Finally, by considering an appropriate factorisation limit, 
the massive particles decouple, and we thus obtain a kinematic algebra and all tree-level BCJ numerators for $D$-dimensional pure Yang-Mills theory. 
\end{abstract} 


\maketitle

\section{Introduction}
Quantum field theory holds many surprising discoveries, one of which is the Bern-Carrasco-Johansson (BCJ) duality between colour and kinematics \cite{Bern:2008qj,Bern:2010ue}. In addition to providing a field-theory underpinning of the Kawai-Lewellen-Tye (KLT)  open-closed string relations \cite{Kawai:1985xq}, the duality hints at a hidden algebraic structure in a variety of gauge theories. Scattering amplitudes in these  theories can be written as a sum of cubic diagrams, each one expressed as the product of a colour and a kinematic factor. The colour factors satisfy Jacobi relations inherited from the gauge-group Lie algebra, and the kinematic numerators satisfy corresponding kinematic Jacobi relations \cite{Bern:2008qj}. Through the double-copy construction, gravitational amplitudes can be obtained from the kinematic numerators. 

A central question is to identify the hidden algebra behind the kinematic relations. In this Letter we provide an explicit construction,
in two related contexts. First we will study the amplitudes in an effective theory of heavy particles coupled to gluons, or gravitons~\cite{Georgi:1990um, Luke:1992cs, Neubert:1993mb, Manohar:2000dt,Damgaard:2019lfh}. 
These theories, which we will refer to as HEFT (heavy-mass effective  field  theory),  
\footnote{Note that in the phenomenological literature, HEFT denotes effective theories involving the BEH boson.}
are obtained from a Yang-Mills (YM) theory, or general relativity, by restricting to the leading-order term in an inverse mass expansion. This is an appropriate approximation for the dynamics of particles with momentum  exchange  much smaller than their masses. Astrophysical black-hole scattering in general relativity satisfies this, and the relevant gravitational amplitudes were recently studied through a gauge-invariant double copy~\cite{Brandhuber:2021kpo}. 
The underlying gauge-theory factors are the central objects, and we will here unravel their algebraic structure, including that of pure YM theory after factorising out the heavy particles. 

The understanding of the kinematic algebra has so far only progressed in small steps. The first successful construction of the algebra was limited to the self-dual sector of YM theory \cite{Monteiro:2011pc}. In that case the algebra corresponds to area-preserving diffeomorphisms, and explicit representations of the generators were found. 
Self-dual YM is far from a complete theory, having vanishing tree amplitudes (apart from a single three-point amplitude for complex momenta)   and a non-CPT invariant spectrum, yet it is the first confirmation of BCJ duality with explicit generators and cubic Feynman rules. 
Another example of the duality was found in the nonlinear sigma model~\cite{Chen:2013fya}, as realised in \cite{Cheung:2016prv} using a cubic Lagrangian. The corresponding kinematic algebra was later~\cite{Cheung:2017yef} tied to that of higher-dimensional Poincar\'{e} symmetry~\cite{Cheung:2017ems}.

Efforts to identify the kinematic algebra have recently been renewed for YM theory~\cite{Chen:2019ywi,Chen:2021chy}, and for HEFT \cite{Brandhuber:2021kpo}. The common idea is to realise the algebra with abstract vector and tensor currents, multiplied through a fusion product. A consistent fusion product was worked out for the maximally-helicity-violating (MHV) and next-to-MHV sectors of YM theory~\cite{Chen:2019ywi,Chen:2021chy}
\footnote{By ${\rm N^{\it k} MHV}$ amplitude in Yang-Mills we mean one whose numerator has the schematic form  $(\varepsilon \Cdot \varepsilon)^{k+1} \prod (\varepsilon \Cdot p)$.}.
The approach was then applied to HEFT in \cite{Brandhuber:2021kpo}, giving explicit expressions for two heavy particles coupled to gluons or gravitons, and the fusion products were presented up to six particles. This approach is well-adapted for gravitational physics, and was used to compute the black hole scattering angle in a post-Minkowskian (PM) expansion at 3PM order~\cite{Brandhuber:2021eyq} (see also~\cite{Bern:2019nnu,Damour:2019lcq,Kalin:2020fhe,DiVecchia:2021bdo,Herrmann:2021tct,Bjerrum-Bohr:2021din,Bern:2021dqo}).

In this Letter we construct a kinematic algebra for HEFT, and by factorisation, infer that the same algebra also works for pure YM theory. In particular, we give a representation of all the generators, and all fusion products needed for computing tree-level HEFT amplitudes with two heavy particles and an arbitrary number of gluons/gravitons.  Interestingly, the obtained fusion product has the same structure as the quasi-shuffle product, known from the mathematical literature, specifically in the context of combinatorial Hopf algebras of shuffles and quasi-shuffles
 \cite{hoffman2000quasi,aguiar2010monoidal,fauvet2017hopf}. 
The quasi-shuffle Hopf algebra generates all ordered partitions for a given set \cite{hoffman2000quasi} (often called $\mathbb{S}\mathbb{C}$ -- the linear species of set compositions, or ordered partitions). 
Mapping the generators to gauge-invariant expressions, we obtain a closed formula for all tree-level BCJ numerators relevant to the HEFT. The numerators are gauge invariant, manifestly crossing symmetric and factorise into lower-point numerators on the massive poles. The underlying quasi-shuffle Hopf algebra implies that the counting of the number of terms in a numerator with $n{-}2$ gluons gives the Fubini number $\mathsf{F}_{n-3}$, which counts the number of ordered partitions of $n{-}3$ elements.

Finally, all the considerations in HEFT directly translate to pure YM theory. The pure-gluon BCJ numerators, and the corresponding expressions for the generators, are obtained from the natural on-shell factorisation limit~\cite{Brandhuber:2021kpo}, which removes the two  heavy particles and replaces them with an  additional gluon (with label $n{-}1$). This is straightforward: replace the heavy-particle velocity $v$ with the last polarisation vector, $v\rightarrow\epsilon_{n-1}$, and impose the last on-shell condition $p^2_{1\ldots n{-}2}\rightarrow 0$.  This operation does not modify the generator fusion rules, and hence YM theory admits the same kinematic algebra. The heavy-mass poles become spurious in this limit, and cancel out once the amplitude is assembled.

\section{The HEFT kinematic algebra}

A novel colour-kinematic duality and double copy for HEFT was obtained in \cite{Brandhuber:2021kpo}, by four of the present authors. Ignoring couplings, the YM and gravity tree amplitudes with two heavy particles and $n{-}2$ gluons/gravitons
are 
\begin{align}
\begin{split}
\label{eq:newDC}
	A(12\ldots n{-}2,v)&\, =\, \sum_{\commut \in \rho} {\npre(\commut,v)\over d_\commut}\, ,
	\\
	M(12\ldots n{-}2,v)&\, =\, \sum_{\commut\in \tilde{\rho}} {\big[\npre(\commut,v)\big]^2 \over d_\commut}\, ,
\end{split}
\end{align}
where $\rho$  ($\tilde{\rho}$) denotes all (un)ordered nested commutators of the particle labels $\{ 1, \ldots \,  , n{-}2 \}$, where the leftmost label is fixed to 1. The ordering is important since  here we work with colour-ordered YM amplitudes.  Considering the set $\{1,2,3\}$, we have    $\rho=\{[[1,2],3],[1,[2,3]]\}$ and $\tilde{\rho}=\{[[1,2],3],[[1,3],2],[1,[2,3]]\}$. 
In general, labels $1, \ldots , n{-}2$ are reserved for the gluons/gravitons and the heavy particles are assigned $n{-}1$ and $n$, and $v$ is the velocity that characterises the heavy particles.

The nested commutators are in one-to-one correspondence with cubic graphs (and hence BCJ numerators), and the corresponding massless scalar-like propagator denominators are denoted as $d_\commut$.  For instance, the nested commutator $[[1,2],3]$ corresponds to the following cubic graph, associated BCJ numerator, and propagator denominator: 
\begin{align}
\begin{tikzpicture}[baseline={([yshift=-0.8ex]current bounding box.center)}]\tikzstyle{every node}=[font=\small]    
   \begin{feynman}
    \vertex (a)[myblob]{};
     \vertex [above=0.3cm of a](b)[dot]{};
     \vertex [left=0.6cm of b](c);
     \vertex [left=0.22cm of b](c23);
     \vertex [above=0.13cm of c23](v23)[dot]{};
    \vertex [above=.4cm of c](j1){$1$};
    \vertex [right=.7cm of j1](j2){$2$};
    \vertex [right=0.5cm of j2](j3){$3$};
   	 \diagram*{(a) -- [thick] (b),(b) -- [thick] (j1),(v23) -- [thick] (j2),(b)--[thick](j3)};
    \end{feynman}  
  \end{tikzpicture}\!\!\!\!\leftrightarrow \,   \npre ({[[1,2],3]},v)   \,, ~~~ 
  d_{[[1,2],3]} = p^2_{12} p^2_{123}\,,
  \end{align}
where  $p_{i_1\ldots i_r}:=p_{i_1}+\cdots+p_{i_r}$ and the red square denotes the heavy-particle source.

The BCJ numerator $\mathcal{N}(\Gamma, v)$ is a function of a nested set of labels $\Gamma$, and it has an expansion which parallels that of the commutator, e.g. 
\begin{align}
    \npre([1,[2,3]],v)&=\,\npre(123,v)-\npre(132,v)\nn\\
    &\,\,-\npre(231,v)+\npre(321,v) \ ,
\end{align}
and we refer to the object
$\npre(1\ldots n{-}2, v)$ as the {\it pre-numerator}. In analogy with a Lie algebra, this quantity should be obtained by multiplying generators through an associative fusion product. Thanks to the nested commutator structure, the BCJ numerators will automatically satisfy kinematic Jacobi identities. 

Explicit pre-numerators can be obtained from the constraint imposed by requiring that they lead to correct amplitudes, and in \cite{Brandhuber:2021kpo} this was done up to six points. 
In  the following, it will be crucial to find representations of the pre-numerators where any non-locality will correspond to a massive physical pole $\sim {1\over v\cdot P}$, where  $P$  is a sum of  gluon  momenta
\footnote{Note that $ \sum_{i=1}^{n-2} v\Cdot p_i=0$. 
}.
This linearised  propagator arises because of the large-mass expansion. Our results will be an improvement compared to \cite{Brandhuber:2021kpo}, since in that work additional spurious poles were present in the pre-numerators. 
We find the following explicit new results up to five points:
\begin{align}
\label{eq:3}
	\npre(1,v)&={v\Cdot \varepsilon_1}\, ,\nn\\
	\npre(12,v)&=-\frac{v\Cdot F_1\Cdot F_2\Cdot v}{2v\Cdot p_1}\, , \nn\\
	\npre(123,v)&=\frac{v\Cdot F_1\Cdot F_2\Cdot F_3\Cdot v}{3v\Cdot p_{1}}-\frac{v\Cdot F_1\Cdot F_2\Cdot V_{12}\Cdot F_3\Cdot v}{3v\Cdot p_1 v\Cdot p_{12}}\nn \\ &
	~~~\,-\frac{v\Cdot F_1\Cdot F_3\Cdot V_1\Cdot F_2\Cdot v}{3v\Cdot p_1 v\Cdot p_{13}}\, ,
\end{align}
where 
 $F^{\mu\nu}_i\!\!:=p^{\mu}_i\varepsilon^{\nu}_i{-}\varepsilon^{\mu}_i p^{\nu}_i$, and 
$V_{\tau}^{\mu \nu}\!\!:= v^\mu \sum_{j\in \tau} p_j^\nu=v^\mu p^{\nu}_\tau$.
Note that gauge invariance is manifest except in the case of $\npre(1,v)$, where it follows from three-point kinematics. 

Following \cite{Chen:2019ywi,Chen:2021chy, Brandhuber:2021kpo}, the pre-numerators are presumed to be constructible in an algebraic fashion, by multiplying abstract generators of the kinematic algebra via a fusion product,
\begin{align}\label{eq:preNumAlg}
	\npre(12\ldots n{-}2, v):=\langle T_{(1)}\star T_{(2)}\star  \cdots \star  T_{(n{-}2)}\rangle\, ,
\end{align}
where the $T_{(i)}$s are generators carrying the gluon label $i$, and $\star$ denotes the bilinear and associative fusion product. 
The angle bracket represents a linear map from the abstract generators to gauge- and Lorentz-invariant functions. It preserves the multi-linearity with respect to the polarisation vectors and the linear scaling in the velocity $v$ of the heavy particles. 

The starting point of the construction is  $\langle T_{(i)}\rangle=v\Cdot \varepsilon_i$, which  is the unique choice that respects  all the properties listed above, and  furthermore generates the correct three-point amplitude. 
 We can then combine two generators to obtain %
\begin{align}
\label{fusionfirst}
	 \npre(12,v):=\langle T_{(1)}\star T_{(2)}\rangle=-\langle T_{(12)}\rangle\, ,
\end{align}
where we choose 
$
\langle T_{(12)}\rangle={v\Cdot F_1\Cdot F_2\Cdot v \over 2v\Cdot p_1}
$ to reproduce Eq.~\eqref{eq:3} 
\footnote{Note that  on the massive pole $
\langle T_{(12)}\rangle$ factorises  as  $\langle T_{(12)}\rangle \big{|}_{v\Cdot p_1 \rightarrow 0} = (-{p_1 \Cdot p_2\over 2 v\Cdot p_1})   \langle T_{(1)}\rangle\langle T_{(2)}\rangle$.}. Similarly, at five points one finds 
\begin{align}
\label{fusionsecond}
	T_{(12)}\star  T_{(3)}&=-T_{(123)}+T_{(12),(3)}+T_{(13),(2)}\, , 
\end{align}
with 
\begin{align}
   &  \langle T_{(123)}\rangle ={v\Cdot F_1\Cdot F_2\Cdot F_3\Cdot v \over 3v\Cdot p_1}\, , \,\,\,
   \langle T_{(12),(3)}\rangle ={v\Cdot F_1\Cdot F_2\Cdot V_{12}\Cdot F_3\Cdot v \over 3v\Cdot p_1 v\Cdot p_{12}}\,, \nn\\
 &   \langle T_{(13),(2)}\rangle={v\Cdot F_1\Cdot F_3\Cdot V_{1}\Cdot F_2\Cdot v \over 3v\Cdot p_1 v\Cdot p_{13}}\, \, .
\end{align}
The particular index assignments in the obtained generators are consistent with a general formula, which we find to work to any number of points,
\begin{align}\label{eq:T}
&\langle T_{(1\tau_1),(\tau_2),\ldots,(\tau_r)}\rangle:=\begin{tikzpicture}[baseline={([yshift=-0.8ex]current bounding box.center)}]\tikzstyle{every node}=[font=\small]    
   \begin{feynman}
    \vertex (a)[myblob]{};
     \vertex[left=0.8cm of a] (a0)[myblob]{};
     \vertex[right=0.8cm of a] (a2)[myblob]{};
      \vertex[right=0.8cm of a2] (a3)[myblob]{};
       \vertex[right=0.8cm of a3] (a4)[myblob]{};
       \vertex[above=0.8cm of a] (b1){$\tau_1~~~~$};
        \vertex[above=0.8cm of a2] (b2){$\tau_2$};
        \vertex[above=0.8cm of a3] (b3){$\cdots$};
         \vertex[above=0.8cm of a4] (b4){$\tau_r$};
         \vertex[above=0.8cm of a0] (b0){$1~~$};
       \vertex [above=0.8cm of a0](j1){$ $};
    \vertex [right=0.2cm of j1](j2){$ $};
    \vertex [right=0.6cm of j2](j3){$ $};
    \vertex [right=0.4cm of j3](j4){$ $};
    \vertex [right=0.8cm of j4](j5){$ $};
      \vertex [right=0.2cm of j5](j6){$ $};
    \vertex [right=0.6cm of j6](j7){$ $};
     \vertex [right=0.0cm of j7](j8){$ $};
    \vertex [right=0.8cm of j8](j9){$ $};
   	 \diagram*{(a)--[very thick](a0),(a)--[very thick](a2),(a2)--[very thick](a3), (a3)--[very thick](a4),(a0) -- [thick] (j1),(a) -- [thick] (j2),(a)--[thick](j3),(a2) -- [thick] (j4),(a2)--[thick](j5),(a4) -- [thick] (j8),(a4)--[thick](j9)};
    \end{feynman}  
  \end{tikzpicture}\nn\\
&={v\Cdot F_{1\tau_1}\Cdot V_{\Theta(\tau_{2})}\Cdot F_{\tau_2}\cdots V_{\Theta(\tau_{r})}\Cdot F_{\tau_r}\Cdot v\over (n{-}2)v\Cdot p_{1}v\Cdot p_{1\tau_1}\cdots v\Cdot p_{1\tau_1\tau_2\cdots \tau_{r-1}}}\, .
\end{align}
The $\tau_i$s are ordered non-empty sets 
such that $\tau_1\cup\tau_2\cup\cdots\cup\tau_r=\{2,3, \ldots ,  n{-}2\}$ and $\tau_i\cap\tau_j=\emptyset$, i.e. they constitute a partition.   
The set $\Theta(\tau_i)$ consists of all indices to the left of $\tau_i$ and smaller than the first index in $\tau_i$; that is $\Theta(\tau_i)= (\{1\}\cup \tau_1 \cup \cdots \cup \tau_{i-1}) \cap \{1,\ldots, \tau_{i[1]}\}$. 
Note that the denominators in Eq.~\eqref{eq:T} are the advertised massive propagators.  For convenience, we also define $F_{\tau_i}$ as the ordered contraction of  several linearised field strengths $F^{\mu\nu}_j$ with indices in $\tau_i$, e.g.~$F^{\mu \nu}_{12} = F^{\mu\alpha}_1 F_{2 \alpha}^{\ \nu}$.  

To clarify the formula, consider a non-trivial example, $T_{(1458),(26),(37)}$, that is mapped to 
\begin{align} \label{eq:ex1}
    \langle T_{(1458),(26),(37)}\rangle={v\Cdot F_{1458}\Cdot V_{1}\Cdot F_{26}\Cdot V_{12}\Cdot F_{37}\Cdot v\over 8v\Cdot p_{1}\, v\Cdot p_{1458}\, v\Cdot p_{124568}} \ . 
\end{align}
We may further clarify the   $\Theta(\tau_i)$s by drawing a ``musical diagram'', where the gluon labels (notes) are filled in progressively from left to right and each horizontal line indicates which set in the partition they belong to:
\begin{align}
\begin{tikzpicture}[baseline={([yshift=-0.8ex]current bounding box.center)}]\tikzstyle{every node}=[font=\small]    
   \begin{feynman}
    \vertex (l1)[]{}; \vertex [left=0.5cm of l1](rm1)[]{$(1\tau_1)$};\vertex [right=4.5cm of l1](r1)[]{}; \vertex [right=0.5cm of l1](v1)[sb]{\white\textbf\small $\mathbf{1}$};\vertex [right=2.0cm of l1](v4)[sb]{\white\textbf\small $\mathbf{4}$};\vertex [right=2.5cm of l1](v5)[sb]{\white\textbf\small $\mathbf{5}$};
    \vertex [right=4.cm of l1](v8)[sb]{\white\textbf\small $\mathbf{8}$};
    \vertex [above=0.3cm of l1](l2)[]{};  \vertex [right=4.5cm of l2](r2)[]{};\vertex [left=0.5cm of l2](rm2)[]{$(\tau_2)$};\vertex [right=1.cm of l2](v2)[sb]{\white\textbf\small $\mathbf{2}$};\vertex [right=3.0cm of l2](v6)[sb]{\white\textbf\small $\mathbf{6}$};
     \vertex [above=0.3cm of l2](l3)[]{};  \vertex [right=4.5cm of l3](r3)[]{};\vertex [left=0.5cm of l3](rm3)[]{$(\tau_3)$};\vertex [right=1.5cm of l3](v3)[sb]{\white\textbf\small $\mathbf{3}$};\vertex [right=3.5cm of l3](v7)[sb]{\white\textbf\small $\mathbf{7}$};
   	 \diagram*{(l1)--[thick](v1)- -[thick](v4)--[thick](v5)--[thick](v8)--[thick](r1),(l2)--[thick](v2)--[thick](v6)--[thick](r2),(l3)--[thick](v3)--[thick](v7)--[thick](r3)};
    \end{feynman}  
  \end{tikzpicture}
\end{align}
A given $\Theta(\tau_i)$ is associated with the first gluon on the horizontal line $\tau_i$, and the set includes  all labels ``southwest'' of this gluon. Specifically, in this example, the relevant sets used in Eq.~\eqref{eq:ex1} are $\Theta(26)=\{1\}$, and $\Theta(37)=\{1,2\}$. Furthermore, the contraction of field strengths can be read out by following each horizontal $\tau_i$-line in this musical diagram. A horizontal line can be thought of as the fundamental representation of the Lorentz group, and the linearised field strengths as Lorentz generators acting in this space.  

Let us return to the algebra of the abstract generators. The pre-numerators can be recursively constructed from only knowing the following fusion product:
\begin{align}
\label{onlyone}
	T_{(1\tau_1),(\tau_2),\ldots,(\tau_r)}\star T_{(j)}.
\end{align}
We assume that the possible outcome of this fusion product maintains the relative order of the labels in the left and right generator. Then by assuming we have a complete set of generators, we can only produce the terms
\begin{align}\label{eq:ansatz}
&T_{(1j),(\tau_1),(\tau_2),\ldots,(\tau_r)},	\quad T_{(1\tau_1),(\tau_2),\ldots,(\tau_i),(j),(\tau_{i+1}),\ldots,(\tau_r)}, \nn\\
& T_{(1\tau_1),(\tau_2),\ldots,(\tau_ij),\ldots,(\tau_r)}, \quad \text{where}\quad i\in \{1,\cdots,r\}.
\end{align}
By writing up a general ansatz, and fixing the free coefficients by comparing to the correct amplitudes via the map in Eq.~\eqref{eq:T}, we find a simple all-multiplicity solution. The fusion product is captured by the general formula  
\begin{align}
\label{fusionone}
	 &T_{(1\tau_1),\ldots, (\tau_r )}\star T_{(j)}= \hskip-0.5cm  \sum_{\sigma\in \{(\tau_1),\ldots, (\tau_{r})\}\shuffle \{(j)\}}T_{(1\sigma_1),\ldots, (\sigma_{r+1})}\nn\\
	 &\hskip2cm-\sum_{i=1}^{r}T_{(1\tau_1),\ldots, (\tau_{i-1}),(\tau_{i}j),(\tau_{i+1}), \ldots,(\tau_{r})}\, ,
\end{align}
where  $\shuffle$ denotes the shuffle product between two sets, e.g.~$\{A, B\}\shuffle \{C\}=\{ABC, ACB,CAB\}.$
A proof for Eq.~\eqref{fusionone} will be given in the next section; here we will study examples. For $n=4, 5$, Eqs.~\eqref{fusionfirst} and   \eqref{fusionsecond} are recovered, and at six points, the fusion products are 
\begin{align}
\label{fusionthird}
	T_{(123)}\star T_{(4)}&=-T_{(1234)}+T_{(123),(4)}+T_{(14),(23)}\nn\\
	T_{(12),(3)}\star T_{(4)}&=-T_{(12),(34)}-T_{(124),(3)}\nn \\ &+T_{(12),(3),(4)}+T_{(12),(4),(3)}+T_{(14),(2),(3)}\nn\\
	T_{(13),(2)}\star T_{(4)}&=-T_{(13),(24)}-T_{(134),(2)}+T_{(13),(2),(4)}\nn \\ & 
	+T_{(13),(4),(2)}+T_{(14),(3),(2)}\, , 
	\end{align}
leading to the six-point pre-numerator
\begin{align}
\label{npre1234}
    \npre(1234,v)&=\langle -T_{ {(12)}, {(3)}, {(4)}}-T_{ {(12)}, {(4)}, {(3)}}-T_{ {(14)}, {(2)}, {(3)}}\nn\\
    &-T_{ {(14)}, {(3)}, {(2)}}-T_{ {(13)}, {(2)}, {(4)}}-T_{ {(13)}, {(4)}, {(2)}}\nn \\
    &+T_{ {(123)}, {(4)}}+T_{ {(124)}, {(3)}}+T_{ {(134)}, {(2)}}\nn\\
    &+T_{ {(12)}, {(34)}}+T_{ {(13)}, {(24)}}+T_{ {(14)}, {(23)}}-T_{ {(1234)}}\rangle\,  .
\end{align}
As already advertised, the algebra defined by the fusion product in Eq.~\eqref{fusionone} 
 is known in the context of combinatorial Hopf algebras of shuffles and quasi-shuffles
 \cite{hoffman2000quasi,aguiar2010monoidal,fauvet2017hopf}. 
Specifically,  our fusion product
defines a  quasi-shuffle Hopf algebra that generates all ordered partitions for a given set \cite{hoffman2000quasi}.  
Indeed, the subscripts of the $T$s are precisely all possible ordered partitions of $\{2,3,\ldots, n{-}2\}$. 
This is also interpreted in \cite{fauvet2017hopf} as 
a Hopf monoid in the category of coalgebra species. 
These  Hopf algebras  are  endowed with a product that is 
commutative and associative  \cite{hoffman2000quasi,cartier2002fonctions,ihara2006derivation}, 
with a  coproduct, counit  and antipode  \cite{hoffman2000quasi} (see the    Appendix   for more details).

We have thus found a realisation of the kinematic algebra for HEFT by mapping it to a quasi-shuffle Hopf algebra. 
Note that the  associativity of the fusion product is a natural property    -- for example, we can construct a BCJ numerator either  as $ ((T_{(1)} \star T_{(2)})\star T_{(3)})\star \cdots $ or   $ \cdots \star (T_{(n-4)} \star (T_{(n{-}3)}\star T_{(n{-}2)}))$. To complete the story, we must also give the fusion product for the most general generators. Assuming the fusion product is associative and preserves the relative order for the left and right generators, one obtains a unique result~\cite{aguiar2010monoidal},
\begin{align}
\label{otherfusion}
	& T_{(1\tau_1),\cdots, (\tau_r )}\star
	T_{(\omega_1),\cdots, (\omega_s )}=\nn\\
	&
	\sum_{\substack{\sigma|_{\{\tau\}}=\{(\tau_1), \cdots,(\tau_{r})\}\\   \sigma|_{\{\omega\}}=\{( \omega_{1}),\cdots,(\omega_s )\}}} (-1)^{t-r-s}T_{(1\sigma_1),(\sigma_2),\cdots, (\sigma_{t})}\, , 
\end{align}
where $\tau_i$ and  $\omega_j$ do not contain the label $1$, as this index is always fixed to be the leftmost index of any expression, and thus it is inert to the algebra. The fusion product of two generators, neither containing label 1, is also given by Eq.~\eqref{otherfusion} after dropping the 1.
We use $\{\tau\}$ and $\{\omega\}$ to denote the total set of labels in the $\tau_i$ and $\omega_i$, respectively. By $\sigma|_{\{\tau\}}$ we mean a restriction to the elements in $\{\tau\}$, 
e.g.~$\{(235),(4),(67)\}|_{\{2,3,4\}}=\{(23),(4)\}$. 

The number of ordered partitions of $\{2,3,\ldots, n{-}2\}$ are known as the Fubini numbers
\cite{mezHo2019combinatorics}
\begin{align}
    \mathsf F_{n{-}3}=\sum_{r=1}^{n{-}3}r!\  {n{-}3\brace r}\, ,
\end{align}
which therefore also counts the number of  terms in the pre-numerator of an $n$-point HEFT amplitude. Here ${n\brace k}$ denotes the number of $k$-partitions on $n$ objects (also known as  Stirling partition number of the second kind). The Fubini numbers also give the Hilbert series of $\mathbb{S}\mathbb{C}$~\cite{fauvet2017hopf}. 

From the kinematic algebra, the closed form of the pre-numerator is directly obtained as
\begin{align}\label{eq:closedForm}
&\npre(1\ldots n{-}2,v)=\sum_{r=1}^{n{-}3}\sum_{\tau\in \mathbf{P}_{\{2,\cdots,n{-}2\}}^{(r)}}({-}1)^{n+r} \langle T_{(1\tau_1),\ldots,(\tau_r)}\rangle\,  ,
\end{align}
where  $\langle T_{(1\tau_1),\ldots,(\tau_r)}\rangle$ is defined in Eq.~\eqref{eq:T} and $\mathbf{P}_{\{2,\cdots,n{-}2\}}^{(r)}$ denotes  all the ordered partitions of  $\{2,3,\ldots, n{-}2\}$ into $r$ subsets. 
This closed-form expression  automatically induces a recursion relation for  the pre-numerator:
\begin{align}
\label{rre}
&\npre(12\ldots n{-}2,v)=\begin{tikzpicture}[baseline={([yshift=-0.8ex]current bounding box.center)}]\tikzstyle{every node}=[font=\small]     \begin{feynman}
    \vertex (a)[myblob]{};
    \vertex[right=0.8cm of a](v2)[myblob]{};
    \vertex[above=0.9cm of v2](jr1){$~~\cdots n{-}2$};
    \vertex[left=0.5cm of jr1](jr2){$2$};
    \vertex [above=0.9cm of a](j1){$1$};
   	 \diagram*{(a) -- [very thick] (v2),(v2) -- [thick] (jr1),(v2) -- [thick] (jr2),(a) -- [thick] (j1)};
    \end{feynman}  
  \end{tikzpicture}+\sum_{\tau_R}\begin{tikzpicture}[baseline={([yshift=-0.8ex]current bounding box.center)}]\tikzstyle{every node}=[font=\small]     \begin{feynman}
    \vertex (a)[myblob]{};
    \vertex[right=0.9cm of a](v2)[myblob]{};
    \vertex[above=0.9cm of v2](jr1){$~~~$};
    \vertex[right=0.3cm of jr1](jr2){\white $0$};
    \vertex[left=0.3cm of jr1](jr3){$~~~~~~\tau_R$};
     \vertex [above=0.3cm of a](b)[dot]{};
      \vertex [above=0.63cm of b](d){$1~~~~\tau_L$};
     \vertex [left=0.6cm of b](c);
     \vertex [left=0.22cm of b](c23);
     \vertex [above=0.13cm of c23](v23)[dot]{};
    \vertex [above=.4cm of c](j1){$~$};
    \vertex [right=0.7cm of j1](j2){$~~$};
    \vertex [right=0.5cm of j2](j3){$~~$};
   	 \diagram*{(a) -- [thick] (b),(a) -- [very thick] (v2),(v2) -- [thick] (jr3),(v2) -- [thick] (jr2),(b) -- [thick] (j1),(v23) -- [thick] (j2),(b)--[thick](j3)};
    \end{feynman}  
  \end{tikzpicture}\nn\\
  &=(-1)^{n}{v\Cdot F_{12\ldots n{-}2}\Cdot v   \over (n{-}2)v\Cdot p_{1}} \\ & -\sum_{\tau_R\subset   \{2,\cdots,n{-}2\}}(-1)^{n_{R}}\npre(1\tau_L,v){(n{-}2{-}n_{R})
  H_{\Theta_{\tau_R}, \tau_R}
  \over (n{-}2)v\Cdot p_{1\tau_L}},\nn
\end{align}
where 
 $\tau_L\cup\tau_R=\{2,3,\cdots, n{-}2\}$,  $\tau_L,\tau_R\neq \emptyset$, and we have defined  \begin{align}
     H_{\sigma, \tau}:=
p_{\sigma}\Cdot F_{\tau}\Cdot v \, . 
\end{align} 
 Here $n_{R}$ denotes the number of indices  in $\tau_R$.
From Eq.~\eqref{rre}, we can see that the number of terms satisfies the recursion relation
\begin{align}
    \mathsf{F}_{n{-}3}=\sum_{i=0}^{n-4}\binom{n{-}3}{i} \mathsf F_{i}\, ,
\end{align}
where $\mathsf{F}_0=1$. This is the well-known  recursion relation for the Fubini numbers \cite{10.2307/2312725}.  To illustrate the simplicity of the pre-numerator, we quote the fairly modest number of terms up to ten points:  
\begin{center}
\begin{tabular}{||c|| c|c| c| c| c| c| c| c 
||} 
 \hline
 $n$ & 3 & 4 & 5 & 6 & 7 & 8 & 9 & 10 
 \\ [0.5ex] 
 \hline
 $\mathsf F_{n{-}3}$  & \,1\, & \,1\,  & \,3\,  & \,13\, & \,75\, & \,541\, & \,4683\, & \,47293\, 
 \, \\ 
 \hline
\end{tabular}
\end{center}
In the next section we  present a general proof of our construction of the BCJ numerators.

\section{Proof of the form of the 
pre-numerator}
\label{sec:factor}

Here we give the proof of the BCJ numerators. Using
$\npre([12{\ldots}n{-}2],v)\!:=\!\npre([{\ldots}[[1,2],3]{\ldots}, n{-}2],v)$, we show that they give correct amplitudes, as obtained from the pre-numerator in Eq.~\eqref{eq:closedForm}. In addition, we will show that the following simple relation holds, valid in the HEFT:  
\begin{align}\label{eq:BCJall}
\npre ([12\ldots n{-}2],v) =
    (n{-}2)\, \npre(12\ldots n{-}2,v) \, .
\end{align}
The outline of the proof is as follows. Starting from 
the factorisation properties  on massive poles  of our HEFT numerators as derived from the KLT formula \cite{Bjerrum-Bohr:2010pnr}, we prove  that the quantity  $(n-2) \npre(12\ldots n{-}2,v)$ has  the same factorisation. We will then consider the difference between $(n-2) \npre(12\ldots n{-}2,v)$ and the  BCJ numerator (as derived from KLT), which is free of  poles. Using arguments similar to those of \cite{Arkani-Hamed:2016rak, Rodina:2016jyz}, we will then show that gauge invariance ensures that this difference vanishes.

The starting point is the  
factorisation on massive poles of BCJ numerators in HEFT.   Using KLT relations, one can easily show that in the on-shell limit $v\Cdot p_{1\tau_L} \rightarrow 0$
\begin{align}\label{eq:factor}
\begin{tikzpicture}[baseline={([yshift=-0.8ex]current bounding box.center)}]\tikzstyle{every node}=[font=\small]    
   \begin{feynman}
    \vertex (a)[myblob]{};
     \vertex [above=0.3cm of a](b)[dot]{};
     \vertex [left=0.6cm of b](c);
     \vertex [left=0.22cm of b](c23);
     \vertex [above=0.13cm of c23](v23)[dot]{};
    \vertex [above=.4cm of c](j1){$1$};
    \vertex [right=.7cm of j1](j2){$2\cdots$};
    \vertex [right=0.5cm of j2](j3){$~~n{-}2$};
   	 \diagram*{(a) -- [thick] (b),(b) -- [thick] (j1),(v23) -- [thick] (j2),(b)--[thick](j3)};
    \end{feynman}  
  \end{tikzpicture} 
  &\begin{tikzpicture}[baseline={([yshift=2.0ex]current bounding box.center)}]\tikzstyle{every node}=[font=\small]    
   \begin{feynman}
    \vertex (a){$\longrightarrow$};
   	 \diagram*{};
    \end{feynman}  
  \end{tikzpicture}
  \begin{tikzpicture}[baseline={([yshift=-0.8ex]current bounding box.center)}]\tikzstyle{every node}=[font=\small]     \begin{feynman}
    \vertex (a)[myblob]{};
    \vertex[right=0.6cm of a](cut1){\red $\bm\times$};
     \vertex [above=0.3cm of a](b)[dot]{};
      \vertex [above=0.6cm of b](d){$1~~~~\tau_L$};
     \vertex [left=0.6cm of b](c);
     \vertex [left=0.22cm of b](c23);
     \vertex [above=0.13cm of c23](v23)[dot]{};
    \vertex [above=.4cm of c](j1){$~$};
    \vertex [right=0.7cm of j1](j2){$~~$};
    \vertex [right=0.5cm of j2](j3){$~~$};
    \vertex[right=1.2cm of a](v2)[myblob]{};
    \vertex [above=0.3cm of v2](bv2)[dot]{};
      \vertex [above=0.6cm of bv2](dv2){$~~~\tau_{R}$};
     \vertex [left=0.6cm of bv2](cv2);
     \vertex [left=0.22cm of bv2](c23v2);
     \vertex [above=0.13cm of c23v2](v23v2)[dot]{};
    \vertex [above=.4cm of cv2](j1v2){$~$};
    \vertex [right=0.7cm of j1v2](j2v2){$~~$};
    \vertex [right=0.5cm of j2v2](j3v2){$~~$};
   	 \diagram*{(a) -- [thick] (b),(a) -- [very thick] (v2),(b) -- [thick] (j1),(v23) -- [thick] (j2),(b)--[thick](j3),(v2) -- [thick] (bv2),(bv2) -- [thick] (j1v2),(v23v2) -- [thick] (j2v2),(bv2)--[thick](j3v2)};
    \end{feynman}  
  \end{tikzpicture} \\
\npre_{\rm KLT}([1\ldots n{-}2], v)& \, {
\longrightarrow} \, \,	p_{_{\Theta(\tau_{R})}}\Cdot p_{\tau_{R[1]}}\,\times\nn\\
&~~~~~\npre_{\rm KLT}([1\tau_L], v) \npre_{\rm KLT}([\tau_R], v)\, , \nn
\end{align}
where 
 $\tau_L \cup \tau_R=\{2, 3,\ldots, n{-}2\}$  and $\tau_{R[1]}$ denote the first index in  $\tau_R$.  We also called  $n_L$ ($n_R$) the number of gluons in $\tau_L$  ($\tau_R$).  The red cross denotes the cut on the physical pole.  The derivation of this formula is given in the Appendix,  also making use of the results of \cite{DelDuca:1999rs,Kleiss:1988ne,Vaman:2010ez,Bjerrum-Bohr:2010diw,Bern:1998sv,Carrasco:2016ldy,Frost:2020eoa}.

The next step is to prove that the pre-numerator in Eq.~\eqref{eq:closedForm} has the same factorisation as Eq.~\eqref{eq:factor}, which we will now do inductively. The seed of the induction is the factorisation where the right-hand side of Eq.~\eqref{eq:factor} only contains one gluon,  that is, $n_{R}{=}1$, and we focus on the 
 massive pole $1\over v\Cdot p_{1\tau_L}$. The factorisation  is then  immediately read off  from Eq.~\eqref{rre}: only one term in the second diagram  in that equation  contributes, with the residue given by 
\begin{align} \label{eq:fact}
\!\!\!  (n{-}3) \, \npre(1i_2\ldots i_{n{-}3},v)\, p_{1i_2\ldots i_{n{-}3}}\Cdot p_{i_{n{-}2}} \, \npre(i_{n{-}2},v)\, .
\end{align}
In the next step, we assume that   factorisation for  $n_{R}=j{-}1$ at the massive pole $1\over v\Cdot p_{1\tau_L}$ has the same form as  Eq.~\eqref{eq:factor}, and we then derive that   for  $n_{R}{=}j$. 
According to Eq.~\eqref{rre}, in this channel the residue of  $(n{-}2)\, \npre$  at the massive pole is 
\begin{align} \label{eq:facgen}
\!\!\! (n_L{+}1)\npre(1\tau_L,v)	 \, \sum_{r=1}^{n_{R}}\sum_{\sigma\in \mathbf{P}_{\tau_R}^{(r)}}{H_{_{\Theta(\sigma_{1})},\sigma_1}\cdots H_{_{\Theta(\sigma_{r})}, \sigma_r}\over v\Cdot p_{\sigma_1}\cdots v\Cdot p_{\sigma_1\sigma_2\cdots \sigma_{r-1}}}\, .
\end{align}
As we show in the  Appendix, in the  limit $v\Cdot p_{1\tau_L} \rightarrow 0$, the  sum in Eq.~\eqref{eq:facgen} becomes precisely a BCJ numerator, 
\begin{align} 
\!\!\! \label{eq:BCJress}
\sum_{r=1}^{n_{R}} \sum_{\sigma\in \mathbf{P}_{\tau_R}^{(r)}}{H_{_{\Theta(\sigma_{1})},\sigma_1}\cdots H_{_{\Theta(\sigma_{r})}, \sigma_r}\over v\Cdot p_{\sigma_1}\cdots v\Cdot p_{\sigma_1\sigma_2\cdots \sigma_{r-1}}}
= p_{_{\Theta(\tau_{R})}}\Cdot p_{\tau_{R[1]}} n_R\, \npre(\tau_R,v)\, .
\end{align}
This establishes that our proposed formula in Eq.~\eqref{eq:closedForm} has the required factorisation property of Eq.~\eqref{eq:factor}. 

Next, we consider the difference
\begin{align}
	f=(n{-}2)\, \npre(1\ldots n{-}2,v)-\npre_{\rm KLT}([1\ldots n{-}2],v).
\end{align}
 As the factorisation  on the heavy-mass poles is the same, and both contain only such poles, $f$ must be  a polynomial. An adaptation of the argument of \cite{Arkani-Hamed:2016rak, Rodina:2016jyz}   allows us to show that $f=0$. To this end, we note that the velocity $v$ can appear in two possible ways. First, through the combination $p\Cdot v$ with $p$ being any of the momenta. This multiplies a polynomial function of dimension $n{-}4$ built from $n{-}2$ gluon momenta and $n{-}2$ polarisations. As is well known, and pointed out recently in \cite{Arkani-Hamed:2016rak, Rodina:2016jyz}, no such gauge-invariant function exists and hence it must vanish. Second, $v$ can appear in the combination $v\Cdot F$, which now multiplies a function of dimensions $n{-}4$, constructed from $n{-}2$ gluon momenta and $n{-}3$ polarisations. As before,  such a  function must vanish. 
 Hence we conclude that $f=0$, and therefore
 \begin{align}
 (n{-}2)\, \npre(1\ldots n{-}2,v)=\npre_{\rm KLT}([1\ldots n{-}2],v)\, .
\end{align}
This completes the  derivation of our  BCJ numerator. As shown in the    Appendix  $\npre_{\rm KLT}$ is crossing symmetric, hence $\npre(1\ldots n{-}2,v)$ has the same property, which leads to Eq.~\eqref{eq:BCJall}. One may also verify Eq.~\eqref{eq:BCJall} explicitly, e.g.~at four points we have 
 $\npre([12],v)=\npre(12,v)-\npre(21,v)={v\Cdot F_1\Cdot F_2\Cdot v \over 2v\Cdot p_1}-{v\Cdot F_2\Cdot F_1\Cdot v \over 2v\Cdot p_2}=2\,\npre(12,v)$. 
 
  \section{From HEFT to Yang-Mills}It is straightforward to obtain the kinematic algebra, and the BCJ numerators, of pure YM theory from the HEFT construction, by exploiting the factorisation property of the HEFT amplitude on a gluon pole  \cite{Brandhuber:2021kpo}. The massive particles decouple on the pole, and we obtain the BCJ numerator for YM amplitudes: 
\begin{align} \label{eq:pureYM}
    \begin{tikzpicture}[baseline={([yshift=-1.5ex]current bounding box.center)}]\tikzstyle{every node}=[font=\small]    
   \begin{feynman}
    \vertex (a)[myblob]{};
    \vertex [above=0.3cm of a](aa){\red{$\bm\times$}};
     \vertex [above=0.5cm of a](b)[dot]{};
     \vertex [left=0.6cm of b](c);
     \vertex [left=0.22cm of b](c23);
     \vertex [above=0.13cm of c23](v23)[dot]{};
    \vertex [above=.4cm of c](j1){$1$};
    \vertex [right=.7cm of j1](j2){$2\cdots$};
    \vertex [right=0.5cm of j2](j3){$~~n{-}2$};
   	 \diagram*{(a) -- [thick] (b),(b) -- [thick] (j1),(v23) -- [thick] (j2),(b)--[thick](j3)};
    \end{feynman}  
  \end{tikzpicture}& \begin{tikzpicture}[baseline={([yshift=2.0ex]current bounding box.center)}]\tikzstyle{every node}=[font=\small]    
   \begin{feynman}
    \vertex (a){$\longrightarrow$};
   	 \diagram*{};
    \end{feynman}  
  \end{tikzpicture}~~~
  \begin{tikzpicture}[baseline={([yshift=-0.8ex]current bounding box.center)}]\tikzstyle{every node}=[font=\small]    
   \begin{feynman}
    \vertex (a){$n{-}1$};
     \vertex [above=0.6cm of a](b)[dot]{};
     \vertex [left=0.6cm of b](c);
     \vertex [left=0.22cm of b](c23);
     \vertex [above=0.13cm of c23](v23)[dot]{};
    \vertex [above=.4cm of c](j1){$1$};
    \vertex [right=.7cm of j1](j2){$2\cdots$};
    \vertex [right=0.5cm of j2](j3){$~~n{-}2$};
   	 \diagram*{(a) -- [thick] (b),(b) -- [thick] (j1),(v23) -- [thick] (j2),(b)--[thick](j3)};
    \end{feynman}  
  \end{tikzpicture}\nn\\
\mathcal{N}^{\rm YM}([1\ldots n{-}1]) &= \mathcal{N}([1\ldots n{-}2], v)\big{|}^{v \rightarrow \epsilon_{n{-}1}}_{ p_{1\ldots n{-}2}^2\rightarrow 0} \, . 
\end{align}
The same replacement should be performed on the expressions of the generators of the algebra given in Eq.~\eqref{eq:T}, with no modification to the fusion rules. 
We have  also explicitly verified Eq.~\eqref{eq:pureYM} for $D$-dimensional YM amplitudes  up to nine points. 

The BCJ numerators thus obtained are manifestly gauge invariant and crossing symmetric for all  gluons except the last  one, $n{-}1$. The price to pay is that 
they also contain  spurious poles  of the form  $\frac{1}{\varepsilon_{n-1}\Cdot P}$, which however can be eliminated in the  complete amplitudes -- in practice, one can use only independent variables (after imposing  on-shell conditions and momentum conservation) in the amplitudes, then terms with  spurious poles should cancel out, or we can simply drop them by hand. 

As for the BCJ numerators, the spurious poles can also be eliminated explicitly. We take the MHV sector of 
the four-point case as an example to illustrate this idea. According to Eqs.~\eqref{eq:closedForm} and \eqref{eq:pureYM}, the corresponding numerator is
\begin{align}
	&\Big(v\Cdot \varepsilon_1 {p_1\Cdot \varepsilon_2 p_{12}\Cdot \varepsilon_3 p_2\Cdot v v\Cdot p_3\over v\Cdot p_1 v\Cdot p_{12}}+	v\Cdot \varepsilon_1 {p_1\Cdot \varepsilon_3 p_{1}\Cdot \varepsilon_2 p_3\Cdot v v\Cdot p_2\over v\Cdot p_1 v\Cdot p_{13}}\nn\\
	&-	v\Cdot \varepsilon_1 {p_1\Cdot \varepsilon_2 p_{2}\Cdot \varepsilon_3 p_3\Cdot v \over v\Cdot p_1}\Big)\Big{|}_{v\rightarrow \varepsilon_{4}}=\varepsilon_1\Cdot \varepsilon_{4} p_1\Cdot \varepsilon_2 p_{12}\Cdot \varepsilon_3\, ,
\end{align} 
which is in agreement with \cite{Chen:2019ywi}. Similarly, we have verified this in the non-MHV sector in several examples.

A natural question arises as to how the known generalised gauge symmetry of the BCJ numerators in YM \cite{Bern:2008qj,Bern:2010yg,Bern:2010ue} 
manifests itself after taking the decoupling limit on the HEFT numerators.  In this limit, the propagator matrix will become degenerate, which implies that one can add or subtract terms in its kernel and obtain a family of valid BCJ numerators. 

Finally, we highlight  potential connections between our and other approaches  in the literature,  e.g.~in \cite{Bridges:2019siz, Mizera:2019blq,He:2021lro, Cheung:2021zvb}. For example, the construction of \cite{Cheung:2021zvb} also maintains gauge invariance and crossing symmetry of $n{-}1$ external legs, and contains linear spurious poles. Intriguingly,  the numerator of \cite{Cheung:2021zvb} for $n$ gluons has $2\, \mathsf F_{n-2}$ terms, while ours has $\mathsf F_{n-2}$. We also note the appearance in \cite{He:2021lro} of Cayley's trees  in  the construction of BCJ numerators, and it is well-known that Fubini numbers are related to such graphs. It would be interesting to explore the connections among  these approaches.

\section{Conclusions}

In this Letter we constructed a kinematic algebra that manifests BCJ colour-kinematics duality in tree-level HEFT and YM theory, and  showed that it can be mapped to a quasi-shuffle Hopf algebra. 
It is intriguing to note that Hopf algebras have already appeared in several different contexts in quantum field theory and string theory,  
e.g.~renormalisation theory \cite{Connes:1999yr}, symbols and co-actions of loop integrals \cite{Goncharov:2010jf,Duhr:2012fh,Abreu:2014cla,Brown:2015fyf}, harmonic sums \cite{Blumlein:2003gb} and string $\alpha'$-expansion \cite{Broedel:2013tta,Fu:2020frx}. 
The obtained kinematic algebra is very simple in terms of the abstract generators, and a non-trivial aspect of the construction is the map between these generators and the kinematic variables (momenta and polarisations), for which we find a simple closed formula that exhibits manifest gauge invariance (see e.g.~\cite{Bjerrum-Bohr:2010pnr,Mafra:2011kj,Du:2017kpo,Edison:2020ehu,Cheung:2021zvb} for other all-multiplicity BCJ constructions).

Several  questions remain open. First, it would be interesting to derive  our BCJ numerators from a Lagrangian description,
which may expose hidden symmetries/structures of the theory. The non-localities of the numerators are both mild and physical in HEFT, thus a Lagrangian approach seems feasible. It may also prove fruitful to try to find representations of the generators in the form of differential operators in kinematic variables, thus re-introducing kinematics in the fusion rules. 
On the mathematical side one may note that a Hopf algebra should have a coproduct and counit: what do these operations imply for the numerator and amplitude? 
The extension of our construction to loop amplitudes, as well as other theories, is an important avenue. Finally, it would be interesting to find a more direct construction of the pure YM kinematic algebra, without passing through the HEFT. We leave these questions for future investigation.

\section*{Acknowledgements}
We thank Kays Haddad, Sanjaye Ramgoolam, Bill Spence and Mao Zeng for useful  discussions, and  Maor Ben-Shahar, Lucia Garozzo, Fei Teng and Tianheng Wang for collaborations on related topics. This work was supported by the Science and Technology Facilities Council (STFC) Consolidated Grants ST/P000754/1 \textit{``String theory, gauge theory \& duality''} and  ST/T000686/1 \textit{``Amplitudes, strings  \& duality''}
and by the European Union's Horizon 2020 research and innovation programme under the Marie Sk\l{}odowska-Curie grant agreement No.~764850 {\it ``\href{https://sagex.org}{SAGEX}''}.
HJ is supported by the Knut and Alice Wallenberg Foundation under grants KAW 2018.0116 ({\it From Scattering Amplitudes to Gravitational Waves}) and KAW 2018.0162, the Swedish Research Council under grant 621-2014-5722, and the Ragnar S\"{o}derberg Foundation (Swedish Foundations' Starting Grant). 
CW is supported by a Royal Society University Research Fellowship No.~UF160350.

\appendix

\section{Appendix: Factorisation of BCJ numerators from KLT relations}

In this section we derive the factorisation properties of the BCJ numerators from KLT. 
We begin by observing that one can define a DDM   \cite{DelDuca:1999rs} basis of amplitudes $A(1\beta,v)$ also for the HEFT theory, where $\beta$ contains all permutations of the gluons $\{2,\ldots, n{-}2\}$. This is possible because of a generalisation of the  Kleiss-Kuijf  (KK) relations \cite{Kleiss:1988ne} which the 
HEFT amplitudes satisfy,  
\begin{align}\label{eq:KK}
	\sum_{\sigma\in \beta_L\shuffle \beta_R}A(\sigma,v)=0\, ,
\end{align} 
where $\beta_L\cup\beta_R=\{1,2,\ldots, n{-}2\}$ and $\beta_{L},\beta_R\neq \emptyset$.  
One can then  extend the results of \cite{Vaman:2010ez}, and  write 
all  colour-ordered HEFT amplitudes in terms of numerators in the DDM basis:   %
\begin{align}
    A(1\beta,v)=\sum_{\alpha\in S_{n{-}3}}\mathsf{m}(1\beta|1\alpha)\, \npre_{\rm KLT}([1\alpha],v)\, .
\end{align}
Thanks to the off-shell condition  $p_{12\ldots n{-}2}^2\neq 0$, the inverse of 
$\mathsf{m}(1\beta|1\alpha)$ exists and is equal to the  standard KLT matrix $\mathcal{S}$ \cite{Bjerrum-Bohr:2010diw}. 
We then get the KLT representation of the BCJ numerator:
\begin{align}\label{eq:BCJNumKLT}
	\npre_{\rm KLT}([1\alpha], v)=\sum_{\beta\in S_{n{-}3}}\mathcal{S}(1\alpha|1\beta) A(1\beta,v)
	\, .
\end{align}
Using the KK relation and the crossing symmetry of the KLT matrix in HEFT, one can show that  $\npre_{\rm KLT}$ is fully crossing symmetric, which leads to 
\begin{align}\label{eq:pre2numAlg0}
    \npre_{\rm KLT}(12\ldots n{-}2,v)
    &={1\over n{-}2}\, \npre_{\rm KLT}([12\ldots n{-}2],v).
\end{align}
Then to prove the closed formula for the pre-numerator  in Eq.~\eqref{eq:closedForm} 
we only need to show that  
\begin{align}\label{eq:pre2numAlg}
    (n{-}2)\, \npre(12\ldots n{-}2,v)=\npre_{\rm KLT}([12\ldots n{-}2],v)\, .
\end{align}
The above equation implies that its left-hand side inherits the   crossing symmetry from   $\npre_{\rm KLT}$, and hence it can be identified as the BCJ numerator  $\npre([12\ldots n{-}2],v)$.

Using Eq.~\eqref{eq:BCJNumKLT}, the known factorisation  of  HEFT amplitudes, and the recursive relation for the KLT matrix~\cite{Bern:1998sv,Bjerrum-Bohr:2010diw,Carrasco:2016ldy,Frost:2020eoa} \begin{align}
    \mathcal{S}(1\ldots j|1\beta_L j\beta_R)=p_{\Theta(j)}\Cdot p_j\, \mathcal{S}(1\ldots j{-}1|1\beta_L \beta_R) \, ,
    \end{align}
    one finally arrives at Eq.~\eqref{eq:factor}.

As an illustration, we consider the five-point case. According to Eq.~\eqref{eq:BCJNumKLT}, we have %
\begin{align}
	\npre_{\rm KLT}([123],v)&=p_{1}\Cdot p_{2}(p_{12}\Cdot p_{3} \,  A(123,v)+p_{1}\Cdot p_{3}A(132,v))\, .\nn
\end{align}
There are three different poles, and the corresponding  factorisations  are %
\begin{center}
\begin{tabular}{||c| c ||} 
 \hline
 pole& residue of $\npre_{\rm KLT}([123],v)$\\
 \hline\hline
 ${1\over v\Cdot p_{12}}$ & $ p_{12}\Cdot p_{3}\, \npre_{\rm KLT}([12],v)\, \npre_{\rm KLT}(3,v)$  \\ [0.5ex] 
 \hline
 ${1\over v\Cdot p_{13}}$  &  $ p_{1}\Cdot p_{2}\, \npre_{\rm KLT}([13],v)\, \npre_{\rm KLT}(2,v)$  \\ [0.5ex] 
 \hline
 ${1\over v\Cdot p_{1}}$  &  $ p_{1}\Cdot p_{2}\, \npre_{\rm KLT}(1,v)\, \npre_{\rm KLT}([23],v)$  \\ [0.5ex] 
 \hline
\end{tabular}
\end{center}
in agreement with Eq.~\eqref{eq:factor}.

\section{Proof of factorisation of BCJ numerators}

In this Appendix we prove Eq.~\eqref{eq:BCJress}. We begin with the factorisation property in Eq.~\eqref{eq:facgen}, which we quote here for convenience, 
\begin{align} \label{eq:facgenApp}
\npre([1\tau_L],v)	 \, \sum_{r=1}^{n_{R}}\sum_{\sigma\in \mathbf{P}_{\tau_R}^{(r)}}{H_{_{\Theta(\sigma_{1})},\sigma_1}\cdots H_{_{\Theta(\sigma_{r})}, \sigma_r}\over v\Cdot p_{\sigma_1}\cdots v\Cdot p_{\sigma_1\sigma_2\cdots \sigma_{r-1}}}\, .
\end{align}
It is convenient to factor out the dependence on the field strength of the added particle, $F_{\tau_{R[1]}}$, which  according to Eq.~\eqref{rre} always appears dotted with $p_{_{\Theta(\tau_{R})}}$. This allows us to rewrite Eq.~\eqref{eq:facgenApp} in the form 
\begin{align}\label{eq:expandFR2}
\npre([1\tau_L],v) \Big[p_{_{\Theta(\tau_{R})}}\Cdot p_{\tau_{R[1]}}\, 
\varepsilon_{\tau_{R[1]}}\Cdot K_R -p_{_{\Theta(\tau_{R})}}\Cdot \varepsilon_{\tau_{R[1]}}\,  p_{\tau_{R[1]}}\Cdot K_R\Big]\, .
\end{align}
The resulting 
$K^{\mu}_R$ has a rather  complex form which will not be needed explicitly in the following. 

The next step is to   prove  that $p_{\tau_{R[1]}}\Cdot K_R=0$.   To do so, we  note that  $p_{\tau_{R[1]}}\Cdot K_R$ has the schematic form  
\begin{align} \label{eq:HHH}
& f_1(p\Cdot v) H_{i_1, \tau_j}+f_2(p\Cdot v) H_{i_1, \tau_{j_1}}H_{i_2, \tau_{j_2}}+\cdots \nn \\ & 
	+ f_{n{-}3}(p\Cdot v) H_{i_1, \tau_{j_1}}\cdots H_{i_{n{-}3}, \tau_{j_{n-4}}}\,  . 
\end{align}
This expression must have degree one in $v$ (or equivalently  in the mass $m$) hence all the $f_i$ with  $i>1$ contain massive poles of the form   $p\Cdot v$ for some $p$.  We will now prove that $f_i=0$ for all $i>1$ by showing that Eq.~\eqref{eq:HHH} cannot
have any massive poles; we will then separately show that terms of the form $f_1 H$ also vanish.

Consider an arbitrary massive pole $p_{\tau_{R_b}}\Cdot v \rightarrow 0$  in the expression in Eq.~\eqref{eq:expandFR2}, as shown in the diagram below,   
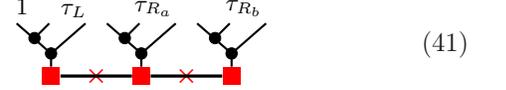
\begin{align}
    \begin{tikzpicture}[baseline={([yshift=-0.8ex]current bounding box.center)}]\tikzstyle{every node}=[font=\small]     \begin{feynman}
    \vertex (a)[myblob]{};
    \vertex[right=0.6cm of a](cut1){\red $\bm{\times}$};
     \vertex [above=0.3cm of a](b)[dot]{};
      \vertex [above=0.6cm of b](d){$1~~~~\tau_L$};
     \vertex [left=0.6cm of b](c);
     \vertex [left=0.22cm of b](c23);
     \vertex [above=0.13cm of c23](v23)[dot]{};
    \vertex [above=.4cm of c](j1){$~$};
    \vertex [right=0.7cm of j1](j2){$~~$};
    \vertex [right=0.5cm of j2](j3){$~~$};
    \vertex[right=1.2cm of a](v2)[myblob]{};
    \vertex[right=0.6cm of v2](cut2){\red $\bm{\times}$};
    \vertex [above=0.3cm of v2](bv2)[dot]{};
      \vertex [above=0.6cm of bv2](dv2){$~~~\tau_{R_a}$};
     \vertex [left=0.6cm of bv2](cv2);
     \vertex [left=0.22cm of bv2](c23v2);
     \vertex [above=0.13cm of c23v2](v23v2)[dot]{};
    \vertex [above=.4cm of cv2](j1v2){$~$};
    \vertex [right=0.7cm of j1v2](j2v2){$~~$};
    \vertex [right=0.5cm of j2v2](j3v2){$~~$};
    \vertex[right=1.2cm of v2](v3)[myblob]{};
    \vertex [above=0.3cm of v3](bv3)[dot]{};
      \vertex [above=0.6cm of bv3](dv3){$~~~\tau_{R_b}$};
     \vertex [left=0.6cm of bv3](cv3);
     \vertex [left=0.22cm of bv3](c23v3);
     \vertex [above=0.13cm of c23v3](v23v3)[dot]{};
    \vertex [above=.4cm of cv3](j1v3){$~$};
    \vertex [right=0.7cm of j1v3](j2v3){$~~$};
    \vertex [right=0.5cm of j2v3](j3v3){$~~$};
   	 \diagram*{(a) -- [thick] (b),(a) -- [very thick] (v2),(v2) -- [very thick] (v3),(b) -- [thick] (j1),(v23) -- [thick] (j2),(b)--[thick](j3),(v2) -- [thick] (bv2),(bv2) -- [thick] (j1v2),(v23v2) -- [thick] (j2v2),(bv2)--[thick](j3v2),(v3) -- [thick] (bv3),(bv3) -- [thick] (j1v3),(v23v3) -- [thick] (j2v3),(bv3)--[thick](j3v3)};
    \end{feynman}  
  \end{tikzpicture}
\end{align}
with  $\tau_{R_a} \cup \tau_{R_b} = \tau_R$.
  This amounts to taking a double residue on the numerator: first $p_{\tau_{R_b}}\Cdot v \rightarrow 0$, then $p_{\tau_{R_a}}\Cdot v \rightarrow 0$. According to the  induction assumption in the derivation in the main text,  
  the result of this operation is 
\begin{align} \label{eq:fff}
&\npre([1\tau_L],v) \,  
p_{_{\Theta(\tau_{R_a})}}\Cdot p_{\tau_{R_a[1]}}
\nn \\ 
\times \,& \npre([\tau_{R_a}],v) p_{_{\Theta(\tau_{R_b})}}\Cdot p_{\tau_{R_b[1]}}\, \npre([\tau_{R_b}],v) \, .
\end{align}
Note that the factor $p_{_{\Theta(\tau_{R_a})}}\Cdot p_{\tau_{R_a[1]}}$, which is arbitrary, is  identical to $p_{_{\Theta(\tau_{R})}}\Cdot p_{\tau_{R[1]}}$  and appears in the first term of Eq.~\eqref{eq:expandFR2}. Hence we  conclude that the second term in that equation  cannot contribute in the limit $p_{\tau_{R_b}}\Cdot v \rightarrow 0$, or equivalently $f_i =0$ for all $i>1$.  

The next step is to  show that the terms of the form  $f_1\, H$ also vanish after performing the sums in Eq.~\eqref{eq:facgenApp}.  This can be verified directly. Collecting all the relevant terms  $T_{\cdots,(\tau_R)}$, $T_{\cdots,(\tau_{R[1]}),(\tau_{R[2]}\cdots \tau_{R[n_R]})}$, $T_{\cdots,(\tau_{R[2]}\cdots \tau_{R[n_R]}),(\tau_{R[1]})}$, we find   
\begin{align}
	&-H_{\tau_{R[1]}, \tau_{R[2]}\ldots \tau_{R[n_R]}}\nn  +H_{\Theta(\tau_{R})\tau_{R[1]}, \tau_{R[2]}\ldots \tau_{R[n_R]}}\nn\\
	&+{v\Cdot p_{\tau_{R[1]}}\over v\Cdot p_{1\tau_L} +v\Cdot p_{\tau_{R[2]}\cdots\tau_{R[n_R]}}}H_{\Theta(\tau_{R}), \tau_{R[2]}\cdots \tau_{R[n_R]}}=0\, ,
\end{align}
where $\tau_{R[i]}$ denotes the $i^{\rm th}$ index in $\tau_R$ and we have used the on-shell condition ${-}v\Cdot p_{\tau_{R[1]}}=v\Cdot p_{1\tau_L} +v\Cdot p_{\tau_{R[2]}\cdots\tau_{R[n_{R}]}}$. This  finishes the proof of  $p_{\tau_{R[1]}}\Cdot K_R=0$. Summarising,  we have shown that Eq.~\eqref{eq:expandFR2} reduces  to
\begin{align}
    \npre([1\tau_L],v) \Big(p_{_{\Theta(\tau_{R})}}\Cdot p_{\tau_{R[1]}}\varepsilon_{\tau_{R[1]}}\Cdot K_R \Big)\, .  
\end{align}
Furthermore, by comparing with Eq.~\eqref{eq:fff}, we see that $\varepsilon_{\tau_{R[1]}}\Cdot K_R$ and $\npre([\tau_R],v)$ have the same factorisation behaviour, and hence their difference  is  a polynomial 
\begin{align}
	g=\varepsilon_{\tau_{R[1]}}\Cdot K_R-\npre([\tau_R],v)\, .
\end{align}
An adaptation of the argument of \cite{Arkani-Hamed:2016rak, Rodina:2016jyz} will now allow us to show that $g=0$. To this end, we note that the velocity $v$ can appear in two possible ways. First, through the combination $p\Cdot v$ with $p$ being any of the momenta. This multiplies a polynomial function of dimension $n_R{-}2$ constructed from $n_R$ gluon momenta and $n_R$ polarisations. According to \cite{Arkani-Hamed:2016rak, Rodina:2016jyz}, no such gauge-invariant function exists and hence it must vanish. Second, $v$ can appear in the combination $v\Cdot F$, which now multiplies a function of dimensions $n_R{-}2$, constructed from $n_R$ gluon momenta and $n_R{-}1$ polarisations. As before,  such a  function must vanish. 
From this  we deduce that $g$  vanishes, ending our proof. 
\section{Appendix: Seven-point pre-numerator}
For completeness, we provide one more example of a pre-numerator, namely at seven points:
\begin{align}
   & \npre(12345,v)=\langle -T_{ {(12)}, {(3)}, {(4)}, {(5)}}-T_{ {(12)}, {(3)}, {(5)}, {(4)}}\nn\\
   & -T_{ {(12)}, {(4)}, {(3)}, {(5)}}-T_{ {(12)}, {(4)}, {(5)}, {(3)}}-T_{ {(12)}, {(5)}, {(3)}, {(4)}}\nn\\
   &-T_{ {(12)}, {(5)}, {(4)}, {(3)}}+T_{ {(12)}, {(3)}, {(45)}}+T_{ {(12)}, {(45)}, {(3)}}\nn\\
   &+T_{ {(12)}, {(34)}, {(5)}}+T_{ {(12)}, {(5)}, {(34)}}-T_{ {(12)}, {(345)}}\nn\\
   &+T_{ {(12)}, {(35)}, {(4)}}+T_{ {(12)}, {(4)}, {(35)}}+T_{ {(123)}, {(4)}, {(5)}}\nn\\
   &+T_{ {(123)}, {(5)}, {(4)}}-T_{ {(123)}, {(45)}}-T_{ {(1234)}, {(5)}}-T_{ {(1235)}, {(4)}}\nn\\
   &+T_{ {(124)}, {(3)}, {(5)}}+T_{ {(124)}, {(5)}, {(3)}}-T_{ {(124)}, {(35)}}-T_{ {(1245)}, {(3)}}\nn\\
   &+T_{ {(125)}, {(3)}, {(4)}}+T_{ {(125)}, {(4)}, {(3)}}-T_{ {(125)}, {(34)}}-T_{ {(13)}, {(2)}, {(4)}, {(5)}}\nn\\
   &-T_{ {(13)}, {(2)}, {(5)}, {(4)}}-T_{ {(13)}, {(4)}, {(2)}, {(5)}}-T_{ {(13)}, {(4)}, {(5)}, {(2)}}\nn\\
   &-T_{ {(13)}, {(5)}, {(2)}, {(4)}}-T_{ {(13)}, {(5)}, {(4)}, {(2)}}+T_{ {(13)}, {(2)}, {(45)}}\nn\\
   &+T_{ {(13)}, {(45)}, {(2)}}+T_{ {(13)}, {(24)}, {(5)}}+T_{ {(13)}, {(5)}, {(24)}}-T_{ {(13)}, {(245)}}\nn\\
   &+T_{ {(13)}, {(25)}, {(4)}}+T_{ {(13)}, {(4)}, {(25)}}+T_{ {(134)}, {(2)}, {(5)}}+T_{ {(134)}, {(5)}, {(2)}}\nn\\
   &-T_{ {(134)}, {(25)}}-T_{ {(1345)}, {(2)}}+T_{ {(135)}, {(2)}, {(4)}}+T_{ {(135)}, {(4)}, {(2)}}\nn\\
   &-T_{ {(135)}, {(24)}}-T_{ {(14)}, {(2)}, {(3)}, {(5)}}-T_{ {(14)}, {(2)}, {(5)}, {(3)}}\nn\\
   &-T_{ {(14)}, {(3)}, {(2)}, {(5)}}-T_{ {(14)}, {(3)}, {(5)}, {(2)}}-T_{ {(14)}, {(5)}, {(2)}, {(3)}}\nn\\
   &-T_{ {(14)}, {(5)}, {(3)}, {(2)}}+T_{ {(14)}, {(2)}, {(35)}}+T_{ {(14)}, {(35)}, {(2)}}\nn\\
   &+T_{ {(14)}, {(23)}, {(5)}}+T_{ {(14)}, {(5)}, {(23)}}-T_{ {(14)}, {(235)}}+T_{ {(14)}, {(25)}, {(3)}}\nn\\
   &+T_{ {(14)}, {(3)}, {(25)}}+T_{ {(145)}, {(2)}, {(3)}}+T_{ {(145)}, {(3)}, {(2)}}-T_{ {(145)}, {(23)}}\nn\\
   &-T_{ {(15)}, {(2)}, {(3)}, {(4)}}-T_{ {(15)}, {(2)}, {(4)}, {(3)}}-T_{ {(15)}, {(3)}, {(2)}, {(4)}}\nn\\
   &-T_{ {(15)}, {(3)}, {(4)}, {(2)}}-T_{ {(15)}, {(4)}, {(2)}, {(3)}}-T_{ {(15)}, {(4)}, {(3)}, {(2)}}\nn\\
   &+T_{ {(15)}, {(2)}, {(34)}}+T_{ {(15)}, {(34)}, {(2)}}+T_{ {(15)}, {(23)}, {(4)}}+T_{ {(15)}, {(4)}, {(23)}}\nn\\
   &-T_{ {(15)}, {(234)}}+T_{ {(15)}, {(24)}, {(3)}}+T_{ {(15)}, {(3)}, {(24)}}+T_{ {(12345)}}\rangle.
\end{align}
As anticipated, it contains 75 terms, which is the Fubini number $\mathsf{F}_{4}$.  For comparison, note that
 the pre-numerators $\npre(1,v)$, $\npre(12,v)$, $\npre(123,v)$ and 
    $\npre(1234,v)$  contain 1, 1, 3 and 13 terms, respectively. 
\section{Appendix: More on   quasi-shuffle Hopf algebras}
As we remarked in the main text,  in a Hopf algebra additional structures are required besides the fusion product. 
 Here  we discuss the  notions of coproduct and  counit (which  make the algebra a bialgebra), and antipode (which make the bialgebra a Hopf algebra).  We begin by recalling that the fusion product is bilinear, that is   $(a T_{\sigma}+bT_{\gamma})\star T_{\rho}=a T_{\sigma}\star T_{\rho}+b T_{\gamma}\star T_{\rho}$, $T_{\rho}\star (a T_{\sigma}+b T_{\gamma}) =a T_{\rho}\star T_{\sigma}+b T_{\rho}\star T_{\gamma}$, where  $a$, $b$ are  numbers. It is also commutative, %
\begin{align}
   T_{(1\tau_1),\cdots, (\tau_r )}\star
	T_{(\omega_1),\cdots, (\omega_s )}=T_{(\omega_1),\cdots, (\omega_s )}\star T_{(1\tau_1),\cdots, (\tau_r )}\, ,
\end{align}
which makes it   an  associative and commutative
quasi-shuffle product. The commutativity is a natural property because the Hopf algebra is associated with  ordered partitions of a given set. This is also consistent with the application to HEFT amplitudes, where we consider the pre-numerator of a particular ordering of gluons, for example the canonical one $\mathcal{N}(12\ldots n{-}2, v)$. Once $\mathcal{N}(12\ldots n{-}2, v)$ is obtained using the fusion products, one can then commute the particle labels and obtain the BCJ numerator $\mathcal{N}([12\ldots n{-}2], v)$.

One can further introduce a number of operations on the generators \cite{hoffman2000quasi}.  
The first one is the  coproduct $\Delta$, which satisfies $\Delta(T_{\sigma})\star \Delta(T_{\gamma})=\Delta(T_{\sigma}\star T_{\gamma})$ and 
\begin{align}
    \Delta(T_{(1\tau_1),\cdots, (\tau_r )})&=T_{(1\tau_1),\cdots, (\tau_r )}\otimes \mathbb{I}-T_{(1)}\otimes T_{(\tau_1),\cdots, (\tau_r )}\nn\\
    &+\sum_{i=1}^{r-1}T_{(1\tau_1),\cdots, (\tau_i )}\otimes T_{(\tau_{i+1}),\cdots, (\tau_r )}\, , \nn\\
    \Delta(T_{(\tau_1),\cdots, (\tau_r )})&=T_{(\tau_1),\cdots, (\tau_r )}\otimes \mathbb{I}+\mathbb{I}\otimes T_{(\tau_1),\cdots, (\tau_r )}\nn\\
    &+\sum_{i=1}^{r-1}T_{(\tau_1),\cdots, (\tau_i )}\otimes T_{(\tau_{i+1}),\cdots, (\tau_r )}\, ,
\end{align}
where $\mathbb{I}$ is the identity element in the algebra with the property $\mathbb{I}\star T_\sigma=T_\sigma\star \mathbb{I}=T_\sigma$. 
The second one is the counit $\epsilon$,  which satisfies 
$\epsilon(T_{\sigma})\star \epsilon(T_{\gamma})=\epsilon(T_{\sigma}\star T_{\gamma})$ and 
\begin{align}
    \epsilon(T_{(\tau_1),\cdots, (\tau_r)})&=0, \quad \epsilon(\mathbb{I})=1\, .
\end{align}
Finally we introduce the antipode $S$ 
, with  the requirement  $\star(\mathbb{I}\otimes S)\Delta(T_{\sigma})=\star( S\otimes \mathbb{I})\Delta(T_{\sigma})=\epsilon(T_{\sigma})\mathbb{I}$, and satisfying  
\begin{align}
    S(T_{(\tau_1),\cdots, (\tau_r)})=&-\sum_{i=1}^{r-1}S(T_{(\tau_1),\cdots, (\tau_i)})\star T_{(\tau_{i+1}),\cdots,(\tau_r)}\nn\, \\
    &- T_{(\tau_{1}),\cdots,(\tau_r)}.
\end{align}

\section{An equivalent kinematic algebra in HEFT and Yang-Mills}

In this section we consider the situation when the gluons are not in  canonical ordering $1,2,\ldots, n{-}2$. 
To do this, we define the kinematic algebra by extending the  generators to contain a superscript accounting for the ordering,  and define the fusion product as   
\begin{align}
\label{eq:Genfusion}
	& T^{(\alpha)}_{(\tau_1),\cdots, (\tau_r )}\star
	T^{(\beta)}_{(\omega_1),\cdots, (\omega_s)}=\nn\\
	&
	\sum_{\substack{\sigma|_{\{\tau\}}=\{(\tau_1), \cdots,(\tau_{r})\}\\   \sigma|_{\{\omega\}}=\{( \omega_{1}),\cdots,(\omega_s )\}}} (-1)^{t-r-s}T^{(\alpha\beta)}_{(\sigma_1),(\sigma_2),\cdots, (\sigma_{t})}\, , 
\end{align}
where the superscripts $\alpha$ and $\beta$ represent the possibly non-canonical orderings of the gluons,   $\alpha$ ($\beta$) is a permutation of $\tau_1\cup\cdots \cup \tau_r $ ($\omega_1\cup\cdots \cup \omega_s $)  and $\alpha \cap \beta = \emptyset$. Furthermore $\tau_i,\omega_j,\sigma_k$ are subsets which preserve the ordering of $\alpha, \beta, \alpha\beta$ respectively.  We  call  $ T^{(\alpha)}_{(\tau_1),\cdots, (\tau_r )}$ the extended generators.

The fusion product of these extended generators is associative but non-commutative, and  can be considered as a non-abelian extension of the quasi-shuffle product. When $\alpha, \beta$ are in  canonical ordering it reduces to the one in the main text, in which case we  omit the generators' superscripts. 
The extended generators can again be mapped to gauge- and Lorentz-invariant functions, which are defined as follows
\begin{align}\label{eq:Tcolor}
&\langle T^{(\alpha)}_{(\tau_1),(\tau_2),\ldots,(\tau_r)}\rangle:=\begin{tikzpicture}[baseline={([yshift=-0.8ex]current bounding box.center)}]\tikzstyle{every node}=[font=\small]       \begin{feynman}
    \vertex (a)[myblob]{};
     \vertex[left=0.8cm of a] (a0)[myblob]{};
     \vertex[right=1.0cm of a] (a2)[myblob]{};
      \vertex[right=0.8cm of a2] (a3)[myblob]{};
       \vertex[right=0.8cm of a3] (a4)[myblob]{};
       \vertex[above=0.8cm of a] (b1){$\overbrace{~~~~~\white {0} ~~}^{\tau_1}$~~~~~~~~~};
        \vertex[above=0.8cm of a2] (b2){$\tau_2$};
        \vertex[above=0.8cm of a3] (b3){$\cdots$};
         \vertex[above=0.8cm of a4] (b4){$\tau_r$};
         \vertex[above=0.8cm of a0] (b0){$~$};
       \vertex [above=0.8cm of a0](j1){$ $};
    \vertex [right=0.2cm of j1](j2){$ $};
    \vertex [right=0.6cm of j2](j3){$ $};
    \vertex [right=0.6cm of j3](j4){$ $};
    \vertex [right=0.8cm of j4](j5){$ $};
      \vertex [right=0.2cm of j5](j6){$ $};
    \vertex [right=0.6cm of j6](j7){$ $};
     \vertex [right=0.0cm of j7](j8){$ $};
    \vertex [right=0.8cm of j8](j9){$ $};
   	 \diagram*{(a)--[very thick](a0),(a)--[very thick](a2),(a2)--[very thick](a3), (a3)--[very thick](a4),(a0) -- [thick] (j1),(a) -- [thick] (j2),(a)--[thick](j3),(a2) -- [thick] (j4),(a2)--[thick](j5),(a4) -- [thick] (j8),(a4)--[thick](j9)};
    \end{feynman}  
  \end{tikzpicture}\nn\\
&={v\Cdot F_{\tau_1}\Cdot V_{\Theta^{\alpha}(\tau_{2})}\Cdot F_{\tau_2}\cdots V_{\Theta^{\alpha}(\tau_{r})}\Cdot F_{\tau_r}\Cdot v\over v\Cdot p_{\tau_{1[1]}}v\Cdot p_{\tau_1}\cdots v\Cdot p_{\tau_1\tau_2\cdots \tau_{r-1}}}\, ,
\end{align}
where $\Theta^{\alpha}(\tau_i)$ denotes the set of all the indices that are to the left  of  the first index of $\tau_i$ in the ordered set $(\alpha)$ and are contained in the subsets $\tau_j$ with $j<i$.  This function vanishes whenever $\Theta^{\alpha}(\tau_i)$ is  the empty set. 

The pre-numerator is then generated from the fusion product of the extended generators through
\begin{align}
    \npre'(i_1i_2\ldots i_{n-2},v)= \langle T^{(i_1)}_{(i_1)}\star T^{(i_2)}_{(i_2)}\star  \cdots \star  T^{(i_{n{-}2})}_{(i_{n{-}2})}\rangle \, .
\end{align}
Importantly, this pre-numerator $\npre'(i_1i_2\ldots i_{n-2},v)$ is  identical to the pre-numerator $\npre(i_1i_2\ldots i_{n-2},v)$   obtained  from an appropriate permutation of the labels in $\npre(12\ldots {n{-}2},v)$
from  the main text. 

We can further define commutation relations using the new fusion product: 
\begin{align}
[T^{(\alpha)}_{\sigma},T^{(\beta)}_{\gamma}]=T^{(\alpha)}_{\sigma}\star T^{(\beta)}_{\gamma} -T^{(\beta)}_{\gamma}\star T^{(\alpha)}_{\sigma}\,  ,
\end{align}
and the extended generators form an infinite-dimensional Lie algebra. 
Then the BCJ numerator for any cubic graph can be expressed as the nested commutator of the corresponding $T^{(j)}_{(j)}$ associated with that graph. 

As an example consider the five-point case, where 
\begin{align}\label{eq:num5c}
   \npre'([[1,2],3],v)&:= \langle [[T^{(1)}_{(1)},T^{(2)}_{(2)}],T^{(3)}_{(3)}]\rangle \\
   &=\langle [T^{(1)}_{(1)}\star T^{(2)}_{(2)}-T^{(2)}_{(2)}\star T^{(1)}_{(1)},T^{(3)}_{(3)}]
   \rangle \, .\nn
\end{align}
We now compute each contribution, starting with 
\begin{align}
& [T^{(1)}_{(1)}\star T^{(2)}_{(2)},T^{(3)}_{(3)}]   =[T^{(12)}_{(1),(2)}{+}T^{(12)}_{(2),(1)}{-}T^{(12)}_{(12)},T^{(3)}_{(3)}]\nn\\
   &=T^{(123)}_{(1),(2),(3)}{+}T^{(123)}_{(1),(3),(2)}{+}T^{(123)}_{(3),(1),(2)}{-}T^{(123)}_{(13),(2)}{-}T^{(123)}_{(1),(23)}\nn\\
   &-(T^{(312)}_{(1),(2),(3)}{+}T^{(312)}_{(1),(3),(2)}{+}T^{(312)}_{(3),(1),(2)}{-}T^{(312)}_{(31),(2)}{-}T^{(312)}_{(1),(32)})\nn\\
  & +T^{(123)}_{(2),(1),(3)}{+}T^{(123)}_{(2),(3),(1)}{+}T^{(123)}_{(3),(2),(1)}{-}T^{(123)}_{(23),(1)}{-}T^{(123)}_{(2),(13)}\nn\\
   & -(T^{(312)}_{(2),(1),(3)}{+}T^{(312)}_{(2),(3),(1)}{+}T^{(312)}_{(3),(2),(1)}{-}T^{(312)}_{(32),(1)}{-}T^{(312)}_{(2),(31)})\nn\\
   &-(T^{(123)}_{(12),(3)}+T^{(123)}_{(3),(12)}-T^{(123)}_{(123)})\nn\\
   &+(T^{(312)}_{(12),(3)}+T^{(312)}_{(3),(12)}-T^{(312)}_{(312)}).
\end{align}
According to the definition in Eq.~\eqref{eq:Tcolor}, all generators with a single index in the first component (i.e.~$T_{(i),\sigma}$) are mapped to zero since a single field strength  $F_i$ is antisymmetric. Furthermore,   $
 T^{(123)}_{(23),(1)}, T^{(312)}_{(12),(3)}$ also vanish due to the fact that the sets $\Theta(1)$ and  $\Theta^{(312)}(3)$ for these two generators are empty. Finally, we arrive at 
\begin{align}
\langle [T^{(1)}_{(1)}\star T^{(2)}_{(2)},T^{(3)}_{(3)}]\rangle&=  \langle T^{(123)}_{(123)} \rangle - \langle T^{(123)}_{(13),(2)} \rangle-\langle T^{(123)}_{(12),(3)} \rangle\nn\\
&-(\langle T^{(312)}_{(312)} \rangle - \langle T^{(312)}_{(31),(2)} \rangle-\langle T^{(312)}_{(32),(1)} \rangle)\nn\\
&=\npre(123,v)-\npre(312,v).
\end{align}
The other term in Eq.~\eqref{eq:num5c} is obtained similarly,  and 
\begin{align}
\langle [T^{(2)}_{(2)}\star T^{(1)}_{(1)},T^{(3)}_{(3)}]\rangle
&=\npre(213,v)-\npre(321,v)\, .
\end{align}
As expected we find
\begin{align}
    \npre'([[1,2],3],v)=\npre([[1,2],3],v)\, .
\end{align}
Therefore, we are able to obtain the BCJ numerators directly from the non-abelian extended quasi-shuffle Hopf algebra. It can be shown for an arbitrary number of gluons that the BCJ numerator defined from the nested commutator of the extended generators is equal to the BCJ numerator obtained from the nested commutator of the indices of the pre-numerator in  the kinematic Hopf algebra. 

Summarising, we can arrive at the BCJ generators in two ways: either via the construction presented in the main text, or with the extended generators introduced here. The diagram below represents these two possibilities, where one can either proceed right and down, or down and right: 
\begin{align}
   & T_{(1)}\star  \cdots \star  T_{(n{-}2)}\xrightarrow[\substack{\text{extended}\\ \text{generators}}]{\text{commutator}} [\ldots[T^{(1)}_{(1)}, T^{(2)}_{(2)}],\ldots, T^{(n-2)}_{(n{-}2)}]\nn\\
   & ~~~~~~~\Big\downarrow\langle ~~\rangle ~~~~~~~~~~~~~~~~~~~~~~~~~~~~ ~~~~~~~~~~~\Big\downarrow \langle ~~\rangle \nn\\
  &  \npre(12\ldots n{-}2,v)~~~\xrightarrow[\text{gluon labels}]{\text{commutator}}~~~~~~\npre([12\ldots n{-}2],v) \, .
\end{align}

\bibliographystyle{apsrev4-1}
\bibliography{ScatEq}
\end{document}